\documentclass[12pt]{article}
\usepackage{epsfig}
\usepackage{amsmath}
\usepackage{bm}
\usepackage[super]{cite}
\usepackage{graphicx}
\begin{document}

\begin{center}
{\bf AB INITIO STUDY OF OPTICAL AND BULK PROPERTIES OF CESIUM LEAD
HALIDE PEROVSKITE SOLID SOLUTIONS}

\vspace{5mm}

{VLADIMIR SALEEV AND ALEXANDRA SHIPILOVA}

{Department of Physics, Samara National Research University,\\
 Moskovskoe Shosse, 34,
Samara, 443086, Russia\\
saleev@samsu.ru} \vspace{5mm}

 \end{center}

\begin{abstract}
 The first principles calculations of band gaps and bulk moduli of cesium
  lead halide perovskite solid solutions, $CsPb(I_{1-x}Cl_x)_3$ and
  $CsPb(I_{1-x}Br_x)_3$, are performed at the level of general gradient
  approximation of the density functional theory. We use supercell
  approach for computational modeling of disordered systems, which
  gives a description of the properties of the structure basing on the
  average over a set of multiple configurations, namely distributions
  of different species over a given set of atomic positions. The
  calculations were performed with the CRYSTAL14 program package.  The
  dependence of the band gap and bulk modulus on the content $x$ are investigated over
  the whole range $0\le x\le 1$.
\end{abstract}


\section{Introduction}
The use of cesium lead halide perovskites, described by the structural
formula $CsPbX_3$ ($X = Cl, Br, I$), as absorbers in solar batteries
has several advantages and disadvantages compared to the hybrid
organic-inorganic perovskites of the $(MA) PbX_3$ or $(FA) PbX_3$
type, where $MA = CH_3NH_3$, $FA = HC(NH_2)_2$. Structures like
$(MA) PbX_3$ have a smaller value of the band gap and the absorption
spectrum for them starts from the visible and infrared range, that is
important since allows the maximal use of the entire spectrum of solar
radiation \cite{1}. On the other hand, such organometallic structures are
optically and thermally unstable, which makes their practical use
difficult. The compound $CsPbI_3$ has a band gap of 1.73~eV, which is
even smaller than that of silicon, widely used as a solar absorber,
while the band gaps for $CsPbBr_3$ and $CsPbCl_3$ are significantly
larger, 2.36 and 3.0~eV \cite{2,3}, accordingly. Thus, $CsPbI_3$ is a good
candidate for the role of an absorber of solar radiation in the
optical and infrared ranges. Recent experimental studies on chemical
and optical properties of the $CsPbI_3$ compound have shown that
partial replacement of iodine by chlorine or bromine stabilizes the
structure and makes such a mixed compound thermodynamically more
stable at room temperatures than the initially pure compound with
iodine \cite{3,4}. It was also observed that the experimentally measured
band gap of such a mixed compound does not increase much as compared
to the pure compound $CsPbI_3$ at the small $x<0.2$.

Here we present a theoretical study for the optical band gap and bulk modulus
dependences in the $CsPb(I_{1-x}Br_x)_3$ and $CsPb(I_{1-x}Cl_x)_3$
cubic phases on the relative content of the iodine and the doping element
(Br or Cl) performed at the level of first-principles quantum mechanical
calculations within the Density Functional Theory (DFT) \cite{5,6}. The
principal difference between doped compounds and pure ones is that
their crystal structure is stoichiometrically disordered and
additional models are necessary to perform DFT calculations of
physical properties. At present, two methods are proposed for
calculations of properties of disordered solid solutions: the supercell
approach \cite{7,8} and the coherent potential approach implemented within
a Green function method \cite{9,10}. The subject of presented study
is a calculation in the supercell approach.

\section{Supercell approach}
Within the supercell approach, periodic conditions are imposed not on
a primitive cell, as in ordered structures, but on a sufficiently
large supercell obtained by scaling the primitive cell. The usual
properties characterizing an ordered structure (cell parameters,
spatial symmetry group, positions of nonequivalent atoms) in this case
refer to the supercell. The size of the supercell should be chosen
such that it corresponds to the content $x$ and reproduces periodical
conditions on the large scale of the supercell. In an ideal crystal,
each atomic position is strictly occupied by an atom of a certain
type, and in a disordered structure, atomic positions are occupied by
atoms of different kinds in a random way, as shown by diffraction
measurements. The idea of the supercell method is that a disordered
structure can be modeled as a statistical ensemble of ordered
structures with all possible configurations of atomic positions. The
statistical method for the theoretical simulation of disordered
compounds is based on obtaining average values of physical quantities
from a set of configurations of strictly ordered compounds, each of
them corresponds to a particular choice for the occupation of atomic
positions in the crystal lattice. Such an approach is justified when
the atoms composing the ordered system are substituted with atoms of
the other constituent in a random way. For a small number of
configurations, averaging over all possible configurations of ordered
compounds can be performed. In the case that the number of possible
configurations is significant, a random sample is used for the average
utilizing the direct Monte Carlo method. Symmetry properties play a
key role in the choice of configurations. The entire set of
configurations is represented as a collection of classes that contain
symmetrically equivalent configurations. The choice of only
symmetrically independent configurations (SIC), by which the averaging
is performed, allows a significant reduction in the number of
configurations during averaging.  All symmetrically independent
configurations can be selected and their multiplicity determined in
each class. {Then, averaging is implemented only by classes of
  independent configurations}, in other words, only for the one
representative of each class. If the number of independent
configurations is large, even after taking into account symmetrically
equivalent configurations, random sampling with weights proportional
to the multiplicities in each symmetric class is carried out.  This
computational scheme is called the symmetric-adapted Monte Carlo
method (SAMC), which makes it possible to calculate structural and
electronic properties of a disordered structure starting from
the average over a finite number of ordered structures \cite{11,12}.

We consider the cubic phase of the compounds $CsPb(I_{1-x}Cl_x)_3$ and
$CsPb(I_{1-x}Br_x)_3$, where $0< x < 1$ in the general case. The
primitive cell contains three positions of halogen atoms, which can be
occupied by iodine or chlorine (bromine) in prescribed proportions,
namely $0: 3$, $1: 2$, $2: 1$, $3: 0$. Obviously, the ordered
structures obtained by multiplication of a primitive cell can not
reproduce the properties of a disordered structure, since the
positions of the atoms are strictly correlated on the scale of one
cell. The minimal supercell that preserves the symmetry properties of
a primitive cubic cell and is large enough to neglect correlations in
the position of atoms is the $2 \times 2 \times 2$ supercell, which
contains 24 positions occupied by halogen atoms.  If $n_I$ is the
number of positions in the simulation cell occupied by iodine atoms,
$n_{Cl}$ is the number of positions occupied by chlorine (bromine)
atoms, the total number of configurations is determined by the number
of combinations equal to
\begin{equation}
N_C=\frac{(n_I+n_{Cl})!}{n_I!n_{Cl}!}
\end{equation}
Taking into account the symmetry properties allows one to pick out
symmetrically independent configurations, which number is much smaller
than the total number of possible configurations. Table 1 shows the
total number of configurations $N_C$ and the number of symmetrically
independent $N_{SIC}$ configurations calculated in the SolidSolution
program \cite{11,12}, which is a part of the CRYSTAL14 \cite{13} software
package, for a $2 \times 2 \times 2$ supercluster with various
proportions between iodine and chlorine or bromine.

\section{Details of calculations}
So, the parameters of the supercell are relaxed until the values at
which the energy minimum of the given structure is reached with an
accuracy of $2\times 10^{-6}$~eV per supercell and the effective
pressure in the cell does not exceed $10^{-2}$~GPa.  With the obtained
relaxed parameters of the supercell, the self-consistent calculation
of the band structure is carried out and the width of the band gap is
guaranteed with an accuracy $3\times 10^{-2}$~eV.

The calculations were carried out with the BECKE exchange functional
\cite{14}, which correctly takes into account the asymptotic of the
exchange interaction at large distances and coincide with the
standard correlation functional PBE \cite{15}. The exchange functional
BECKE effectively takes into account the contribution of the
Hartree-Fock exchange interaction, although it is not a hybrid
potential in the strict sense. As it was shown in \cite{14}, its use allows
to reproduce the values of the band gap for a large number of
dielectrics, which are usually obtained using hybrid potentials, since
calculations with exchange potentials in the density functional
theory, without taking into account the Hartree-Fock {contribution},
strongly underestimate the value of the band gap.  Calculations using
hybrid potentials significantly increase the computation time, which
is a fundamental difficulty in the method of supercells where it is
necessary to average over a large number of configurations.

In the package CRYSTAL14, a mixed numerical-analytical method for
calculation of the overlap integrals is realized for this purpose; the
wave functions of the electrons are sought in the basis of Gaussian
atomic orbitals.  For the number of electrons in an atom $Z <35$,
there are all-electron basic sets. However, for arbitrary $Z$, the
effective-core potential method can be used, when only valence
electrons are explicitly taken into account, and the potential of the
nucleus and internal electrons is replaced by an effective potential
of the "atomic core". Calculations with the all-electron basis are much
more expensive in terms of processor time than calculations with an
effective atomic-core potential, and since we need to perform multiple
computations of the energy of a supercell of 40 atoms, the effective
core pseudo-potentials were used not only for many-electron atoms like
lead ($Z = 82$), cesium ($Z = 55$) and iodine ($Z = 53$), but also for
bromine ($Z = 35$) and chlorine ($Z = 17$).

Electron configuration of lead is described by the effective
pseudopotential PbHAYWLC-211(1d)G with four-electronic valence
basis, cesium - CsSCHAYWSC-31G with two-electronic basis, iodine -
IHAYWLC-31G with two-electronic basis, bromine - BrHAYWLC-31 with
two-electronic basis, chlorine - ClLCHAYWLC -31G with two-electronic
basis.  The effective atomic potentials of the type HAYWLC (large
core) or HAYWSC (small core) are proposed in \citen{16} and have
been repeatedly tested in calculations of a different type, both for
molecules and crystals. {The parameters} of the electronic bases of
lead and cesium were used with the standard values given in the
library on the website of the CRYSTAL14 program. The values of the
diffusion exponential index for iodine, bromine and chlorine were
changed to {achieve a} best description of the energy band gaps of
pure lead-halogen compounds: $CsPbI_3$, $CsPbBr_3$ and $CsPbCl_3$.
Used parameters of basis sets for cesium, lead, iodine, bromine and
chlorine are collected in Appendix. In Fig.~\ref{figure:1}, the
electronic band structures for pure compounds are shown: orange --
$CsPbI_3$, red -- $CsPbBr_3$, blue -- $CsPbCl_3$. The band gaps
{calculated in the} point of high symmetry R agree with data with an
error of about 0.05~eV.

\begin{figure}[th]
    \centerline{\includegraphics[width=14pc]{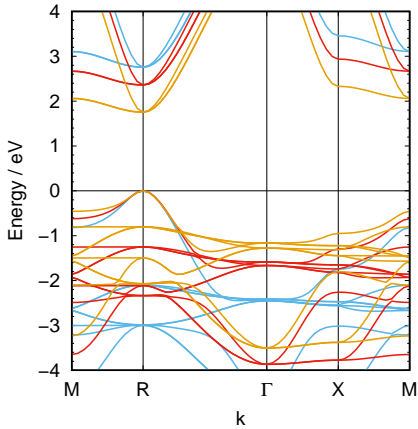}}\vspace*{8pt}%
    \caption{\label{figure:1}The
            electronic band structures for pure compounds: orange --
            $CsPbI_3$, red -- $CsPbBr_3$, blue -- $CsPbCl_3$.}
\end{figure}

\section{Dependence of band gaps on dopant content}

The width of the band gap for each class of symmetrically independent
atomic configurations of the ordered structure $E_{gap}^l$  can be
calculated within the DFT. The width of the band gap of a disordered
structure, corresponding to a certain proportion of halogens in the
composition, is determined by the formula
\begin{equation}
E_{gap}=\sum_{l=1}^{L}w_lE_{gap}^l,
\label{formula:1}
\end{equation}
where $w_L=\displaystyle\frac{n_L}{\sum_{l=1}^L n_l}$, $L=N_{SIC}$, and $n_L$ is
the number of configurations of a certain symmetry class. For cases
where $N_{SIC}\le 200$ the averaging was carried out for all
symmetrically independent configurations. The standard deviation is
calculated by the formula: $D(E_{gap})=(E_{gap}^2-(E_{gap})^2)^{1/2}$,
where $E_{gap}^2=\sum_{l=1}^{L}w_l(E_{gap}^l)^2$. With $N_{SIC}>200$,
the average was performed over 200 randomly chosen configurations,
i.e. in the formula \ref{formula:1} it was assumed that
$N_{SIC}=200$. At each given ratio between halogens, a set of
symmetrically independent ordered structures is generated in the first
step and the multiplicity of $n_L$ configurations in each set is
determined.

\begin{table}[pt]
    \caption{\label{ex} Band gaps as a function of doping
        content $x$}
                  \begin{center}
                  {\begin{tabular}{@{}ccccc@{}}  Content $x$, &  &
    &
 $E_{gap}$, eV  &$E_{gap}$, eV \\
 Cl or Br &  $N_{C}$ &  $N_{SIC}$ & $CsPb(I_{1-x}Cl_x)_3$ &
  $CsPb(I_{1-x}Br_x)_3$ \\
                  0&1&1&1.75&1.75\\
                  1/6&343&80&$1.93\pm0.13$&$1.90\pm0.12$\\
                  1/3&21252&2664&$2.07\pm0.17$&$1.98\pm0.13$\\
                  1/2&87980&8797&$2.19\pm0.14$&$2.03\pm0.14$\\
                  2/3&21252&2664&$2.34\pm0.17$&$2.06\pm0.13$\\
                  5/6&343&80&$2.54\pm0.20$&$2.16\pm0.04$\\
                  1&1&1&$2.90$&$2.36$\\
        \end{tabular}}
        \end{center}
\end{table}

\begin{figure}[th]
\centerline{\includegraphics[width=5cm]{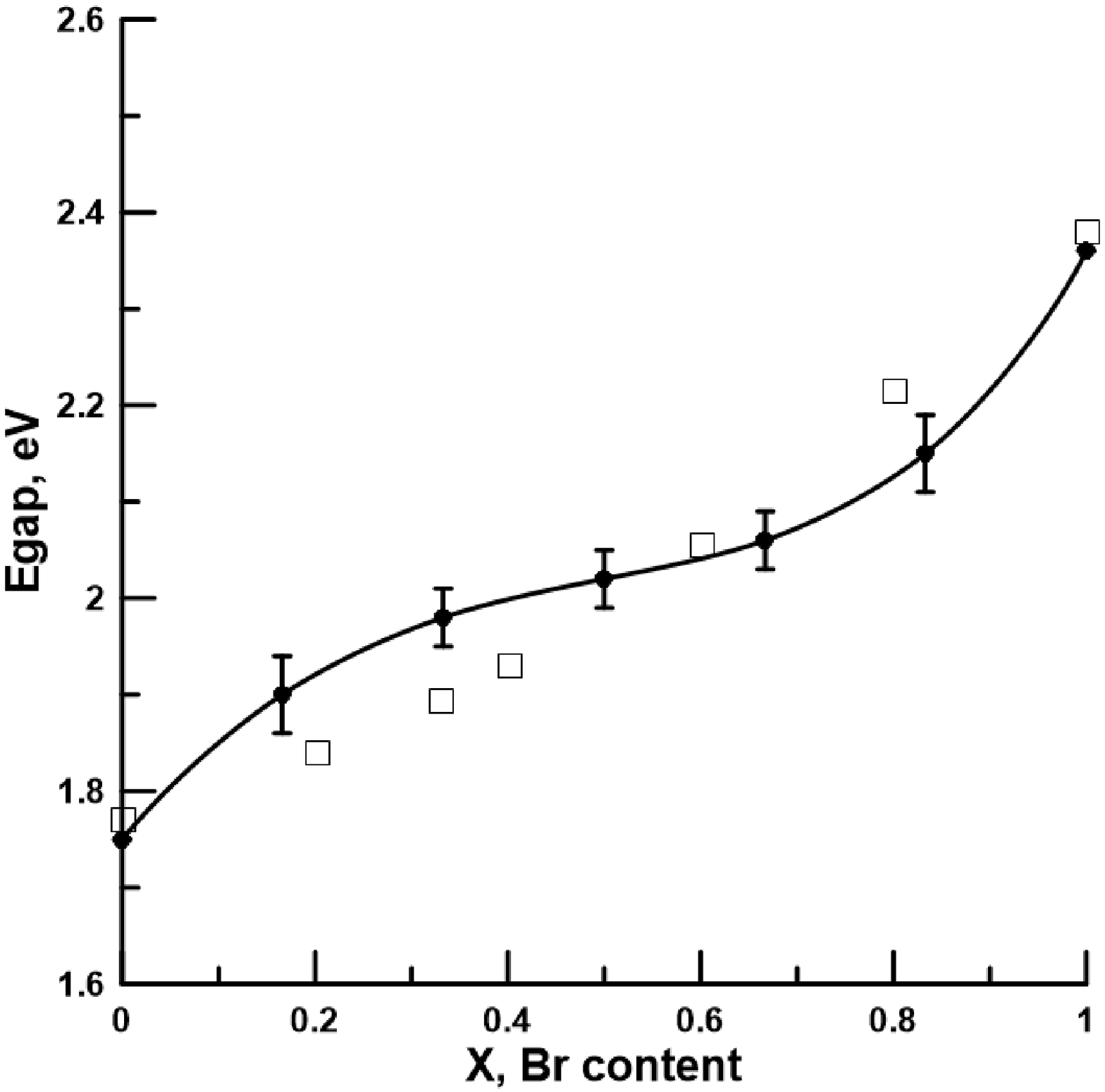}
\includegraphics[width=5cm]{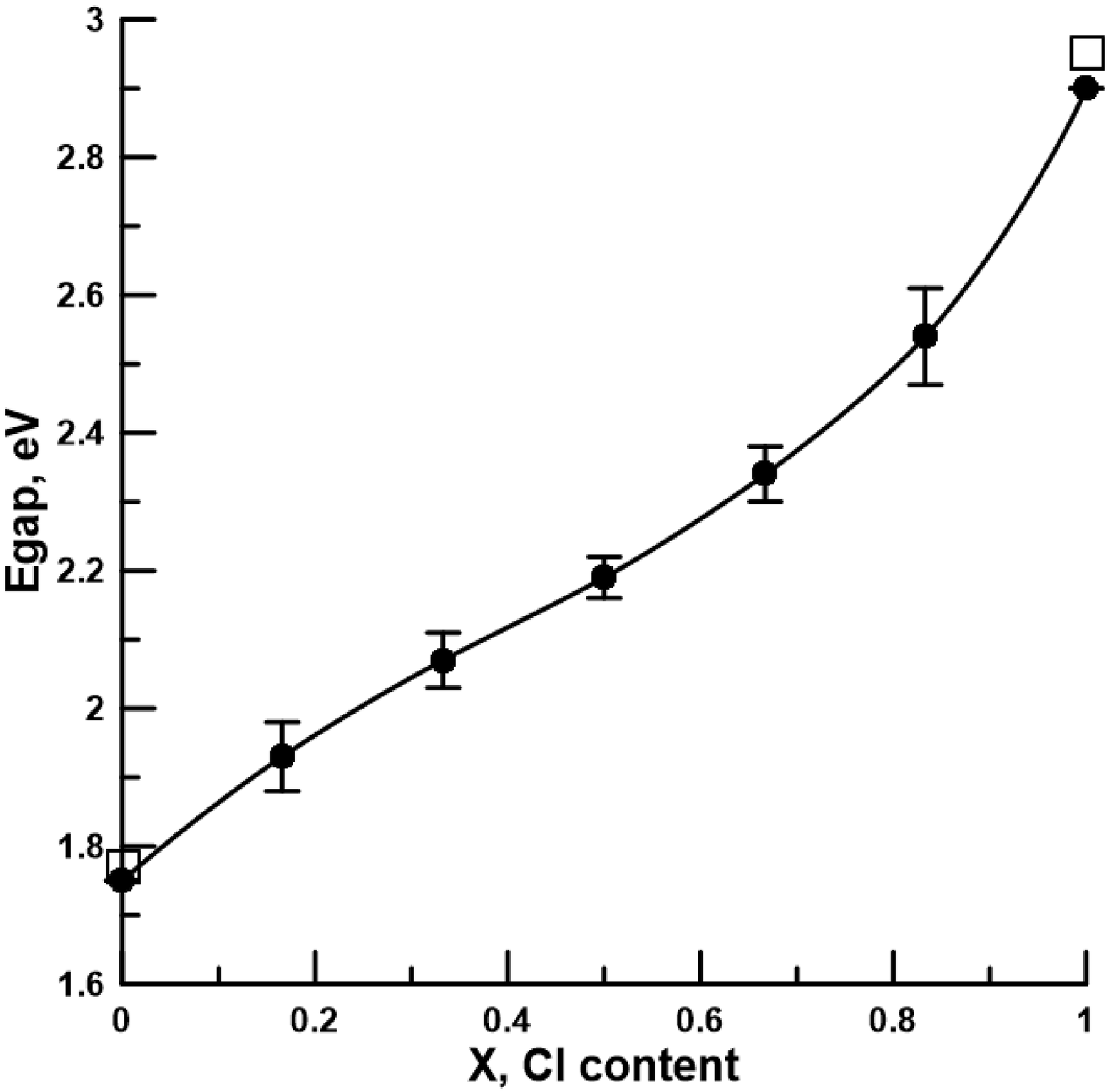}}
\vspace*{8pt}
\caption{Band gap energy for $CsPb(I_{1-x}Br_x)_3$ (left panel) and $CsPb(I_{1-x}Cl_x)_3$ (right panel) as a function of doping content, $x$. Boxes on the left and right panels are the experimental data from \citen{2,3}. Error bars correspond statistical errors $\Delta E_{gap}$.\label{figure:2}}
\end{figure}

The results of {computation of} the band gaps as a function of doping
content $x$ are collected in {the} Table \ref{ex} (columns 4 and 5) and in {the} Fig.~\ref{figure:2}.
The calculated dependence of the band gap on the impurity content deviates
substantially from the linear dependence for the $CsPb(I_{1-x}Br_x)_3$
compound, as well as the available experimental data \cite{3}. Last ones
are described fairly well by a linear law with the different slope for
regions $0<x<0.5$ and $0.5<x<1$. The calculated dependence on $x$ for
the $CsPb(I_{1-x}Cl_x)_3$  compound is in a  good agreement with the
expected linear law, but experimental data for this compound are
absent yet and our calculations can be considered as predictions.

\begin{figure}[th]
\centerline{\includegraphics[width=14pc]{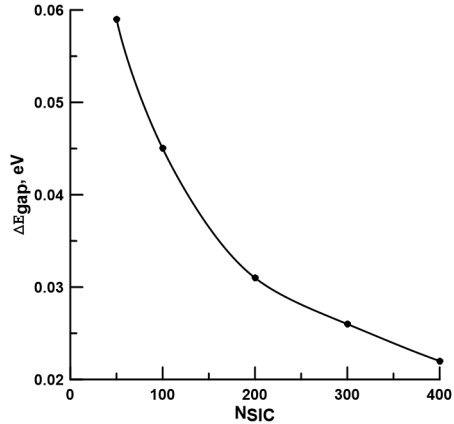}}
\vspace*{8pt}
\caption{\label{figure:3}Statistical error $\Delta E_{gap}$ for band gap energy for structure $CsPb(I_{1-x}Br_x)_3$ with $x=0.5$ as function of number of symmetrically independent configurations, $N_{SIC}$.}
\end{figure}

To demonstrate the convergence of MC procedures, we show in
Fig.\ref{figure:3}  the statistical error for band gap energy for
structure $CsPb(I_{1-x}Br_x)_3$ with $x=0.5$ as a function of number of
symmetrically independent configurations, $N_{SIC}$. The statistical
error is defined as follows: $\Delta E_{gap}\simeq 3
D(E_{gap})/\sqrt{N_{SIC}}$. Our choice, $N_{SIC}=200$,
guarantees an error about 0.03~eV for average values of band gaps.

\section{Dependence of bulk moduli on dopant content}

The {computation scheme of bulk moduli $x$-dependence} for discussed here solid solutions is the same as for {the} $x$-dependence of band gaps. We calculate bulk moduli for {a certain} number of symmetrically independent ordered configurations using
the third-order Birch-Murnaghan isothermal equation of state (based on the Eulerian strain)\cite{17}. {The latter reads}:
\begin{equation}
E(V)=E_0+\frac{9 V_0 B}{16}\left\{\left[ \left(\frac{V_0}{V}\right)^{2/3}-1\right]^3 B'+\left[ \left(\frac{V_0}{V}\right)^{2/3}-1 \right]^2\left[  6-4 \left(\frac{V_0}{V}\right)^{2/3}\right]  \right\},
\label{birch}
\end{equation}
where $B$ is the equilibrium bulk modulus, $B'$ is the first derivative of $B$ with respect to the pressure,  $V_0$ and $E_0$ are the equilibrium volume and energy, at zero pressure. To find $B$, we should obtain the $E(V)$ curve.  Taking into account the huge number of configurations, we need to restrict a number of points in the $E(V)$ curve to the lowest feasible value. Thereby, we use five-points {fitting} procedure with variation coefficient from 0.98 {to} 1.02 for volumes around the equilibrium volume $V_0$ {which} corresponding variation coefficient equal to $1$.

\begin{table}[pt]
    \caption{\label{ex2} Bulk moduli as a function of dopant
        content $x$}
        \begin{center}
{\begin{tabular}{@{}ccccc@{}}
                  Content $x$, &   &
    &
 $B$, GPa  &$B$, GPa \\
  Cl or Br &  $N_{C}$ &  $N_{SIC}$ & $CsPb(I_{1-x}Cl_x)_3$ & $CsPb(I_{1-x}Br_x)_3$ \\
                  0&1&1&16.15&16.15\\
                  1/6&343&80&$13.23\pm0.34$&$14.09\pm0.23$\\
                  1/3&21252&2664&$12.97\pm0.29$&$14.20\pm0.13$\\
                  1/2&87980&8797&$13.16\pm0.35$&$14.84\pm0.18$\\
                  2/3&21252&2664&$14.27\pm0.39$&$15.86\pm0.20$\\
                  5/6&343&80&$17.04\pm0.23$&$17.45\pm0.21$\\
                  1&1&1&$24.37$&$19.17$\\

        \end{tabular}}
        \end{center}
\end{table}

\begin{figure}[h]
\centerline{
        \includegraphics[width=14pc,angle=270]{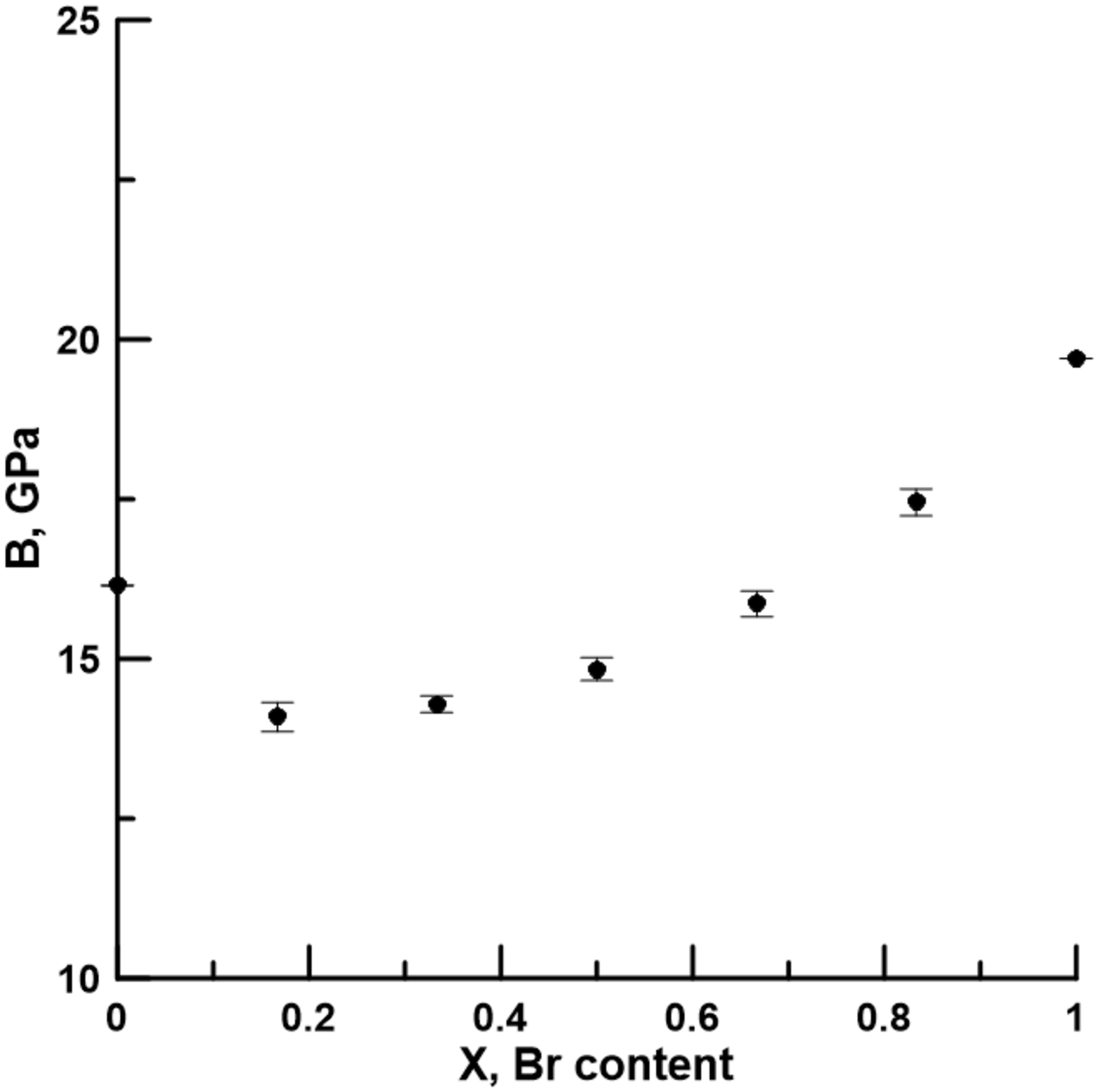}
        \includegraphics[width=14pc,angle=270]{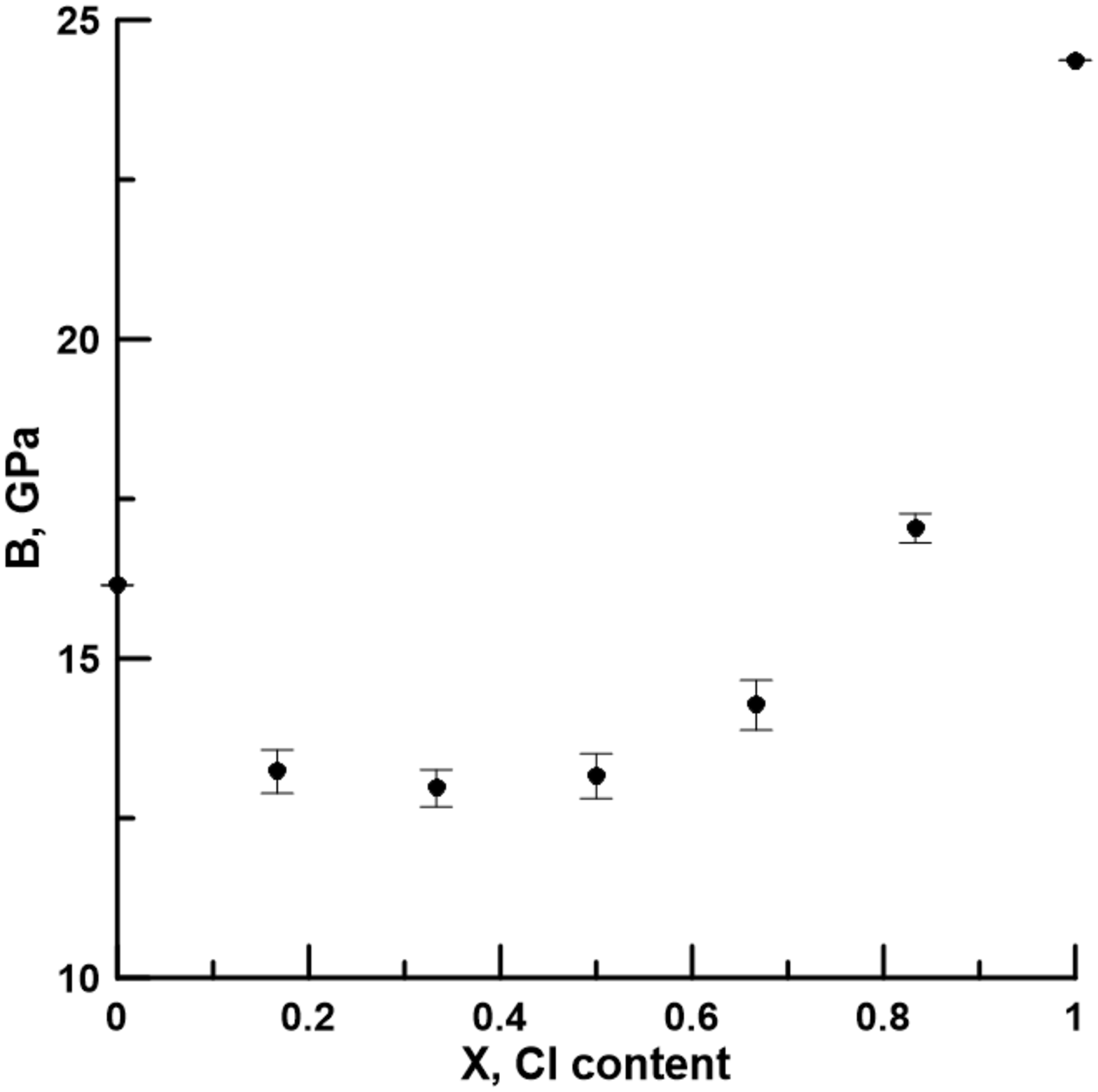}}
    \vspace*{8pt}
    \caption{\label{figure:4}Bulk moduli for $CsPb(I_{1-x}Br_x)_3$ (left panel) and $CsPb(I_{1-x}Cl_x)_3$ (right panel) as function of dopant content, $x$. Error bars correspond statistical errors $\Delta B$. }
\end{figure}

The results of our calculations for bulk moduli are presented in {the} Table \ref{ex2} and Fig.\ref{figure:4}. Bulk moduli of pure $PbCsI_3$ and $PbCsCl_3$ are equal to 16.15 GPa and 24.37 GPa, correspondingly. We should {expect} linear growth of bulk moduli with chlorine content $x$, from 0 to 1, taking into account that chemical bonding for chlorine atoms is stronger than for iodine atoms. {But in reality}, we {obtain a fall} of bulk moduli {with chlorine doping of $PbCsI_3$ till the minimum value of} B=12.97 GPa at the $x=1/3$. Only at the $x=5/6$ bulk modulus for solid solution $CsPb(I_{1-x}Cl_x)_3$ becomes equal to the bulk modulus of pure crystal $PbCsI_3$. The same picture is realized {also} for solid solution $CsPb(I_{1-x}Br_x)_3$: at the $x=1/6$ one has $B=14.09$ GPa {despite} $B=16.15$ for pure $PbCsI_3$. The obtained result seems non-trivial and should be experimentally {examined}.

\section{Conclusions}
In this work, we {performed}  the first-principles calculations for band gap energies, $E_{gap}$, and bulk moduli, $B$, as functions of
the relative dopant amount ($x$) of bromine and chlorine for cubic phases
of cesium {lead iodide} perovskite $CsPbI_3$.  Ab initio calculations for
solid solutions were performed in the supercell approach. In case of band gap dependence, we have
obtained an approximate agreement with experimental data for cubic
phase of $CsPb(I_{1-x}Br_x)_3$ and we have predicted $x$-dependence
for cubic phase of $CsPb(I_{1-x}Cl_x)_3$ at the whole range $0\le x\le
1$. Our predictions for bulk modulus $x$-dependencies show the strong non-linear behaviour, which demonstrates {the}  importance of {disordering}  effects over the chemical content for the cesium lead-halide solid solutions.

\section{Acknowledgments}
The authors thank the Ministry of Education and Science of the Russian
Federation for financial support in the framework of the Samara
University Competitiveness Improvement Program among the world's
leading research and educational centers for 2013-2020, the task
number 3.5093.2017/8.9. Arthur Ernst and Victor Yushankhai are acknowledged for useful
discussions.

\section*{Appendix}

\begin{verbatim}
Basis set for chlorine:
217 2
HAYWLC
0 1 3 7. 1.
4.221 -0.0338 -0.0657
1.769 -0.3131 0.090
0.498 .8138 0.5776
0 1 1 0. 1.
0.131 1.0 1.0

Basis set for bromine:
235 2
HAYWLC
0 1 3 8. 1.
1.547 -0.3827 -0.3476
1.164 0.0798 0.3994
0.320 0.9091 0.5937
0 1 1 0. 1.
0.092 1.0 1.0

Basis set for iodine:
253 2
HAYWLC
0 1  3  7.  1.
1.520    0.5606  -0.0300
1.252   -1.0108  -0.0719
0.295    0.9297   0.6278
0 1 1 0. 1.
0.083    1.0       1.0

Basis set for cesium:
255 2
HAYWSC
0 1  3  8.  1.
0.986  -1.1924  -0.8737
0.833   0.8133   0.9449
0.330   0.9076   0.5185
0 1  1   1. 1.
0.147   1.0      1.0

Basis set for lead:
282 4
HAYWLC
0 1 2 4. 1.
1.335104   -0.1448789   -0.1070612
0.7516086  1.0          1.0
0 1 1 0. 1.
0.5536686  1.0          1.0
0 1 1 0. 1.
0.1420315  1.0          1.0
0 3 1 0. 1.
0.1933887  1.0
\end{verbatim}

\end{document}